\documentclass[11pt]{article}
\usepackage[utf8]{inputenc}
\usepackage[T1]{fontenc}
\usepackage{lmodern}
\usepackage{amsmath,amssymb}
\usepackage{booktabs}
\usepackage{hyperref}
\usepackage[margin=1in]{geometry}
\usepackage{listings}
\title{Partial Partial Aggregates}
\author{Claude Brisson\\\texttt{claude.brisson@gmail.com}}
\date{}

\begin{document}
\maketitle

\begin{abstract}
We introduce partial partial aggregates (PPA), a query optimization technique for distributed engines that pushes only the local compute phase of an aggregate operation through joins. A query that aggregates after a join involves two logical operations, each requiring a network shuffle. Pushing a full aggregate (COMPUTE$\rightarrow$DISTRIBUTE$\rightarrow$MERGE) below the join introduces a third shuffle. In the specific case where the join key is included in the grouping key and the join is FK-PK, the full pushed aggregate can eliminate the top-level aggregate entirely, making it the preferred choice. In all other key configurations, the top aggregate must remain, and the extra shuffle is wasteful. A PPA pushes only COMPUTE, achieving data reduction before the join without the extra shuffle. The technique relies on the distributive property of aggregates and requires accurate NDV estimation~\cite{ndv} for cost-based decisions.
\end{abstract}

\section{Introduction}

Consider a query that aggregates after a join. In a distributed engine, this involves two logical operations: the join and the aggregate. Each requires a network shuffle--the join shuffles by join key (or broadcasts the smaller side), the aggregate shuffles by grouping key. That is two shuffles in the baseline case.

Aggregate pushdown~\cite{yan} reduces the join's input by pushing an aggregate below it. But in a distributed engine, pushing a full aggregate means pushing COMPUTE$\rightarrow$DISTRIBUTE$\rightarrow$MERGE. The pushed DISTRIBUTE introduces a \emph{third} shuffle: the pushed aggregate's shuffle, the join's shuffle, and the top aggregate's shuffle.

This third shuffle is justified only if it enables eliminating the top aggregate. When the join is a FK-PK equijoin and the join key is included in the grouping key, the pushed aggregate produces the final grouping directly--the join adds dimension columns but no duplicates. Full pushdown (PA) then reduces the total to two shuffles by removing the top aggregate. PA is the right choice here.

In all other cases--disjoint keys, partial overlap, non-FK joins--the top aggregate must remain, and PA means three shuffles. Partial partial aggregates (PPA) avoid this: push only COMPUTE (no DISTRIBUTE, no MERGE), achieving data reduction before the join while keeping the shuffle count at two.

Section~\ref{sec:background} reviews distributed aggregation and the extra shuffle problem. Section~\ref{sec:keys} analyzes the key relationships that determine whether PA or PPA is preferred. Section~\ref{sec:ppa} introduces the PPA operator. Section~\ref{sec:cost} discusses the cost model and a decision tree visualization. Section~\ref{sec:physical} details physical operator strategies. Section~\ref{sec:related} positions PPA relative to prior work.

\section{Background}
\label{sec:background}

\subsection{Distributed Aggregation}

In a distributed engine, a logical aggregate \texttt{AGG(GROUP BY key)} becomes three physical operators~\cite{theseus}:

\begin{verbatim}
COMPUTE  ->  DISTRIBUTE  ->  MERGE
(local)      (by key)        (combine)
\end{verbatim}

\begin{itemize}
\item \textbf{COMPUTE}: accumulates values into a local hash table, outputting (key, accumulator) pairs.
\item \textbf{DISTRIBUTE}: shuffles data by grouping key so that all rows with the same key reach the same node.
\item \textbf{MERGE}: combines accumulators from different nodes into the final result.
\end{itemize}

This decomposition requires that aggregation functions be distributive ($\text{SUM}$, $\text{COUNT}$, $\text{MIN}$, $\text{MAX}$) or rewritable in distributive terms ($\text{AVG} \rightarrow \text{SUM}/\text{COUNT}$).

\subsection{Aggregate Pushdown Through Joins}

Pushing an aggregate below a join reduces the join's input volume~\cite{yan}. To traverse a join, the pushed aggregate may need additional grouping keys to preserve join semantics. Consider:

\begin{lstlisting}
SELECT category, SUM(amount)
FROM orders JOIN products ON orders.product_id = products.id
GROUP BY category
\end{lstlisting}

Since \texttt{category} comes from \texttt{products}, aggregating \texttt{orders} alone requires adding the join key \texttt{product\_id} to the grouping set. The pushed aggregate groups by \texttt{(category, product\_id)}, and a top-level aggregate reduces to the final \texttt{GROUP BY category} after the join.

\subsection{Column Equivalence via Join Predicates}
\label{sec:equivalence}

An equijoin predicate \texttt{R.a = S.b} establishes column equivalence: any reference to \texttt{S.b} in the grouping set can be substituted with \texttt{R.a}, and vice versa. Combined with functional dependencies from FK-PK constraints (a primary key determines all columns of its table), this substitution can move dimension-side grouping columns to the fact side.

For instance, if \texttt{products.id} is the primary key, then \texttt{id} $\rightarrow$ \texttt{category}. Via the equijoin \texttt{products.id = orders.product\_id}, grouping by \texttt{product\_id} on the fact side is a refinement of grouping by \texttt{category}. The optimizer can push \texttt{GROUP BY product\_id} to the \texttt{orders} side, then reduce to \texttt{GROUP BY category} after the join.

\subsection{The Extra Shuffle}
\label{sec:extra-shuffle}

Without pushdown, two shuffles occur: the join's and the aggregate's DISTRIBUTE. When a full aggregate (PA) is pushed below the join, a third shuffle is introduced:

\begin{samepage}
\begin{verbatim}
MERGE (GROUP BY category)                -- top aggregate
  DISTRIBUTE (by category)               -- shuffle #3
    COMPUTE (GROUP BY category)
      PHYSICAL_JOIN                      -- shuffle #2
        MERGE (GROUP BY category, product_id)
          DISTRIBUTE (by product_id)     -- shuffle #1 (extra)
            COMPUTE (GROUP BY category, product_id)
              orders
        products
\end{verbatim}
\end{samepage}

Shuffle~\#1 exists only to finalize the pushed aggregate. Whether this extra cost is justified depends on whether the top aggregate can be eliminated.

\section{Key Relationships and Pushdown Strategy}
\label{sec:keys}

Let $g$ denote the set of grouping columns and $j$ the join key columns (on the fact side, after column equivalence substitution per Section~\ref{sec:equivalence}).

\subsection{When PA Eliminates the Top Aggregate}

When the join is a FK-PK equijoin with the aggregate pushed to the fact side, and $j \subseteq g$ (the join key is included in the grouping key), the pushed aggregate produces one row per group with the join key preserved. The join maps each group to exactly one dimension row--no duplicates are introduced. The top aggregate becomes unnecessary and can be removed.

In this specific case, PA reduces the plan to two shuffles (the pushed aggregate's DISTRIBUTE and the join) and is generally the preferred strategy. The decision remains cost-based: the COMPUTE overhead and the pushed DISTRIBUTE's cost must be offset by the data reduction.

\subsection{When PA Adds an Extra Shuffle}

In all other key configurations--$g \subset j$, $j \cap g = \emptyset$, or partial overlap--the top aggregate cannot be eliminated:

\begin{itemize}
\item If $g \subset j$, the pushed aggregate groups at finer granularity than the final result; a reducing top aggregate is needed.
\item If $j \cap g = \emptyset$, the pushed aggregate must add the join key to its grouping set, producing groups strictly finer than the final result.
\item In partial overlap cases, the top aggregate similarly remains.
\end{itemize}

With the top aggregate in place, PA introduces a third shuffle. PPA (Section~\ref{sec:ppa}) avoids it--but whether PPA is worthwhile remains a cost-based decision. COMPUTE reduces data only when the grouping keys have low local cardinality within each processing unit. On sorted or pseudo-sorted columns, each batch contains mostly unique values for the grouping key, and COMPUTE's volume reduction becomes negligible. The NDV estimates from the companion paper~\cite{ndv} inform this decision.

\section{Partial Partial Aggregates}
\label{sec:ppa}

\subsection{Definition}

A \textbf{partial partial aggregate} (PPA) is a logical aggregate operator that maps to COMPUTE only--no DISTRIBUTE, no MERGE. It performs local accumulation and outputs (key, accumulator) pairs, but does not finalize the aggregation across nodes.

\subsection{Eliminating the Extra Shuffle}

Replacing the full pushed aggregate with a PPA removes shuffle~\#1:

\begin{verbatim}
MERGE (GROUP BY category)                -- top aggregate
  DISTRIBUTE (by category)               -- shuffle #2
    COMPUTE (GROUP BY category)
      PHYSICAL_JOIN                      -- shuffle #1
        COMPUTE (GROUP BY category, product_id)  -- PPA
          orders
        products
\end{verbatim}

The PPA (bottom COMPUTE) reduces the \texttt{orders} side before the join. The join may produce duplicates on \texttt{(category, product\_id)}. The post-join COMPUTE re-aggregates. A single DISTRIBUTE$\rightarrow$MERGE completes the aggregation. The shuffle count is the same as no pushdown--two--but the join processes fewer rows.

\subsection{Correctness}
\label{sec:correctness}

The correctness of PPA rests on the distributive property of the aggregate functions:

\begin{enumerate}
\item The pushed COMPUTE outputs one row per distinct grouping key.
\item The join may fan out these rows (one-to-many or many-to-many).
\item The post-join COMPUTE re-aggregates, absorbing duplicates.
\item MERGE combines partial results across nodes.
\end{enumerate}

For distributive aggregates, $\text{SUM}(a, b, c) = \text{SUM}(\text{SUM}(a, b), c)$: the intermediate COMPUTE boundaries are transparent to the final result. The join does not need to preserve aggregate semantics--only the final MERGE must produce the correct answer. Duplicates are absorbed, not prevented.

\subsection{When PPA Is Beneficial}
\label{sec:when}

PPA is beneficial when the top aggregate cannot be eliminated ($j \not\subseteq g$) and the COMPUTE reduction ratio is favorable:

\begin{equation}
\text{reduction ratio} = \frac{\text{ndv}(\text{grouping keys})}{\text{input rows}}
\label{eq:reduction}
\end{equation}

PPA is not beneficial when grouping keys have high cardinality (COMPUTE produces as many rows as it consumes) or when the build side is small enough for a broadcast join regardless.

\section{Cost Model}
\label{sec:cost}

\subsection{Decision Framework}

The optimizer evaluates three strategies for each aggregate-above-join pattern:

\begin{enumerate}
\item \textbf{No pushdown}: aggregate entirely after the join. Two shuffles.
\item \textbf{PA}: push COMPUTE$\rightarrow$DISTRIBUTE$\rightarrow$MERGE below the join. Two shuffles if top aggregate eliminated ($j \subseteq g$, FK-PK), three otherwise.
\item \textbf{PPA}: push only COMPUTE below the join. Two shuffles. Top aggregate always remains.
\end{enumerate}

PA is preferred when it eliminates the top aggregate; PPA is preferred when the top aggregate must stay. The decision is always cost-based: even when PA \emph{can} eliminate the top aggregate, the COMPUTE overhead might not be justified if the reduction ratio is poor.

\begin{equation}
\text{push COMPUTE if: } \text{ndv}(\text{grouping keys}) < \text{input rows} \times \theta
\end{equation}

where $\theta$ is a threshold accounting for COMPUTE overhead (hash table construction, memory pressure).

\subsection{NDV Estimation}

Accurate NDV of the grouping keys is critical for this decision. Options include HyperLogLog sketches (requiring writer-side storage), sampling (requiring data access), or various heuristics--sometimes just a hard-coded constant per data type. The companion paper~\cite{ndv} introduces a metadata-based approach at zero cost, deriving NDV from Parquet dictionary sizes and row group min/max statistics.

\subsection{Batch-Level Estimation}
\label{sec:batch}

The companion paper provides a \emph{global} NDV estimate for the entire column. But COMPUTE operates on batches--rowsets of a fixed size $B$. The optimizer must predict COMPUTE's output volume per batch.

Using the coupon collector model~\cite{feller}, a batch of $B$ rows drawn from a population of $\text{ndv}_{\text{global}}$ distinct values contains an expected:
\begin{equation}
\text{ndv}_{\text{batch}} = \text{ndv}_{\text{global}} \times (1 - e^{-B / \text{ndv}_{\text{global}}})
\label{eq:batch-ndv}
\end{equation}

This $\text{ndv}_{\text{batch}}$ is the output cardinality of COMPUTE: the reduction ratio per batch is $\text{ndv}_{\text{batch}} / B$.

For scan operators, $B$ is typically large. However, reducing operators--filters, joins, aggregates--can shrink $B$ substantially. In practice, batch size management is entrusted to I/O operators: DISTRIBUTE (for aggregates) and EXCHANGE (for joins) aggregate smaller batches back to an efficient size~\cite{theseus}.

The coupon collector model assumes well-spread data--each batch sees a representative sample of distinct values. For sorted or pseudo-sorted columns, each batch contains a localized value range, $\text{ndv}_{\text{batch}} \approx B$, and COMPUTE's volume reduction becomes negligible. The distribution detection from the companion paper~\cite{ndv} identifies this condition.

\subsection{Pushdown Decision Tree}
\label{sec:decision-tree}

Our implementation relied on Calcite's VolcanoPlanner, which explores a search space of alternative plans and selects the cheapest. The result is a tree where each node represents a \emph{choice} among alternative physical implementations.

To visualize this search space, we introduced a compact notation based on standard indented plan representations. The leading two-space indentation at each level is replaced with numbered alternatives: \texttt{1.}~\texttt{2.}~\texttt{3.}\ldots\ The chosen path is marked with \texttt{>} instead of \texttt{.} at each decision point. Each line's suffix summarizes row count and memory cost, enabling quick comparison of alternatives.

Consider a query where $j \subseteq g$:

\begin{lstlisting}
SELECT product_id, SUM(amount)
FROM orders JOIN products ON orders.product_id = products.id
GROUP BY product_id
\end{lstlisting}

\begin{verbatim}
1. No pushdown                                  10K rows   200KB
1.   AGG(product_id, SUM(amount))               10K rows   200KB
1.     JOIN                                      1M rows   170MB
1.       SCAN(orders)                            1M rows    80MB
1.       SCAN(products)                         10K rows     1MB
2> PA / AGG eliminated                          10K rows   1.2MB
2>   JOIN                                       10K rows   1.2MB
2>     MERGE(product_id)                        10K rows   200KB
2>       DISTRIBUTE(product_id)                100K rows     2MB
2>         COMPUTE(product_id)                 100K rows     2MB
2>           SCAN(orders)                        1M rows    80MB
2>     SCAN(products)                           10K rows     1MB
3. PPA / AGG kept                               10K rows   200KB
3.   AGG(product_id, SUM(amount))               10K rows    12MB
3.     JOIN                                    100K rows    12MB
3.       COMPUTE(product_id)                   100K rows     2MB
3.         SCAN(orders)                          1M rows    80MB
3.       SCAN(products)                         10K rows     1MB
\end{verbatim}

\noindent PROJECT operators are omitted for clarity. Here $j = g = \{\texttt{product\_id}\}$ and the join is FK-PK, so the COMPUTE output is bounded by $|\texttt{products}| = 10\text{K}$ rows. Three root-level alternatives are evaluated. PA (option~2, marked~\texttt{>}) eliminates the top aggregate entirely. PPA (option~3) achieves the same pre-join data reduction but retains the top aggregate, making it costlier than PA in this configuration.

For the earlier running example (\texttt{GROUP BY category}), $j \not\subseteq g$ and option~2 is unavailable--PPA would then be preferred over both no pushdown and PA (which would add an extra shuffle without eliminating the top aggregate).

Further choices down the tree--join strategy (broadcast vs.\ shuffle), scan implementation, index usage--are nested in the same way, each level presenting numbered alternatives with cost suffixes. The full tree captures the optimizer's entire search space in a single readable text representation.

Conceptually, nothing prevents pushing COMPUTE further down, inside the scan operator itself. Some engines already do this: Spark's aggregate pushdown into Parquet scans~\cite{spark-pushdown} performs partial aggregation during I/O, reducing data volume before it enters the execution pipeline.

\section{Physical Operators}
\label{sec:physical}

\subsection{Join Strategies}

A logical \texttt{JOIN} maps to a physical join implemented as either:

\begin{itemize}
\item \textbf{Broadcast}: the smaller (build) side is replicated to all nodes; the larger (probe) side stays in place.
\item \textbf{Shuffle}: both sides are redistributed by the join key.
\end{itemize}

\begin{verbatim}
BROADCAST_JOIN:             SHUFFLE_JOIN:

PHYSICAL_JOIN               PHYSICAL_JOIN
  |-- probe                   |-- EXCHANGE (by key)
  +-- BROADCAST               |     +-- probe
        +-- build             +-- EXCHANGE (by key)
                                    +-- build
\end{verbatim}

Adaptive execution (e.g., Spark AQE) can convert a shuffle join to a broadcast join at runtime when actual sizes are small enough. With accurate NDV estimation~\cite{ndv}, this decision can often be made at planning time.

\subsection{Three Strategies Compared}
\label{sec:comparison}

The running example yields three distinct physical plans.

\paragraph{No pushdown.} The aggregate runs entirely after the join:

\begin{verbatim}
MERGE (GROUP BY category)
  DISTRIBUTE (by category)
    COMPUTE (GROUP BY category)
      PHYSICAL_JOIN
        orders
        products
\end{verbatim}

\paragraph{Full pushdown (PA).} A complete aggregate is pushed below the join:

\begin{verbatim}
MERGE (GROUP BY category)
  DISTRIBUTE (by category)
    COMPUTE (GROUP BY category)
      PHYSICAL_JOIN
        MERGE (GROUP BY category, product_id)
          DISTRIBUTE (by product_id)
            COMPUTE (GROUP BY category, product_id)
              orders
        products
\end{verbatim}

\paragraph{PPA.} Only COMPUTE is pushed:

\begin{verbatim}
MERGE (GROUP BY category)
  DISTRIBUTE (by category)
    COMPUTE (GROUP BY category)
      PHYSICAL_JOIN
        COMPUTE (GROUP BY category, product_id)
          orders
        products
\end{verbatim}

PPA achieves the same pre-join data reduction as full pushdown, without the extra DISTRIBUTE$\rightarrow$MERGE cycle.

\section{Related Work}
\label{sec:related}

\paragraph{Eager and lazy aggregation.} Yan and Larson~\cite{yan} formalize pushing aggregates through joins, distinguishing eager aggregation (push below the join) from lazy aggregation (defer above the join). Their framework considers pushing or deferring the full aggregate. PPA introduces a third option: pushing only the local computation phase, a decomposition not considered in the original work.

\paragraph{Partial preaggregation.} Larson and Guo~\cite{larson-preagg} describe partial preaggregation as a local operator that tolerates incomplete aggregation--multiple records per group are acceptable, and final aggregation merges them. Their concern is memory efficiency on a single machine: when the hash table grows too large, partial results are flushed. PPA addresses a different problem: in distributed systems, separating COMPUTE from DISTRIBUTE to avoid an extra shuffle. The ``partial'' in PPA refers to pushing only the compute phase of a distributed aggregate, not to incomplete local accumulation.

\paragraph{Adaptive query processing.} Deshpande et al.~\cite{deshpande} survey techniques for adapting query plans at runtime. PPA naturally enables adaptation: by deferring DISTRIBUTE, the engine can observe actual COMPUTE output sizes before choosing between broadcast and shuffle for the join.

\paragraph{GPU query processing.} Theseus~\cite{theseus} implements aggregate pushdown below joins in a GPU-accelerated distributed query engine. Its memory-based cost model favors transformations that reduce memory footprint--even at the cost of additional computation--making PPA particularly attractive: reducing data volume before GPU hash joins directly reduces GPU memory pressure.

\section{Conclusion}

Partial partial aggregates address a specific gap in distributed aggregate pushdown: when the top-level aggregate cannot be eliminated, pushing a full aggregate introduces an extra network shuffle. PPA avoids this by pushing only COMPUTE through the join, achieving data reduction at the same shuffle cost as no pushdown.

The key relationships between grouping and join keys determine which strategy is preferred. When the join key is included in the grouping key and the join is FK-PK, PA can eliminate the top aggregate and should be preferred. In all other configurations, PPA avoids the extra shuffle while delivering the same pre-join data reduction.

Correctness follows from the distributive property of aggregation: duplicates introduced by joins are absorbed by subsequent COMPUTE and MERGE phases. The technique was deployed in Theseus, where it contributed to reducing data movement in GPU-accelerated distributed query processing. Cost-based decisions require accurate NDV estimation, addressed in the companion paper~\cite{ndv}.

\end{document}